\documentclass{jpsj-suppl}
\usepackage{txfonts,lineno} 
\newcommand{\Xm}{$X_{\rm{max}}\,$}

\title{Report from the Multi-Messenger Working Group at UHECR-2014 Conference}

\author{
Timo \textsc{Karg}$^{1}$ for the IceCube Collaboration$^{6}$\\ 
Jaime \textsc{Alvarez-Mu\~niz}$^{2}$, Daniel \textsc{Kuempel}$^{3}$ and Mariangela \textsc{Settimo}$^{4}$ for the Pierre Auger Collaboration$^{7}$\\ 
Grigory \textsc{Rubtsov}$^{5}$ and Sergey \textsc{Troitsky}$^{5}$ for the Telescope Array Collaboration$^{8}$
}

\inst{
$^{1}$DESY, Platanenallee 6, Zeuthen, Germany\\
$^{2}$Universidad de Santiago de Compostela, Santiago de Compostela, Spain\\
$^{3}$III.\ Physikalisches Institut A, RWTH Aachen University, Sommerfeldstra\ss e, Aachen, Germany\\
$^{4}$Laboratoire de Physique Nucl\'{e}aire et de Hautes Energies (LPNHE), Universit\'{e}s Paris 6 et Paris 7, CNRS-IN2P3, Paris, France\\
$^{5}$Institute for Nuclear Research of the Russian Academy of Sciences, Moscow, Russia\\
$^{6}$https://icecube.wisc.edu/collaboration/authors/2015/04 \\
$^{7}$http://www.auger.org/archive/authors\_2015\_04.html \\
$^{8}$http://www.telescopearray.org/index.php/research/collaborators
}

\email{timo.karg@desy.de, jaime.alvarezmuniz@gmail.com, kuempel@physik.rwth-aachen.de, mariangela.settimo@lpnhe.in2p3.fr, grisha@ms2.inr.ac.ru, sergey.troitsky@gmail.com}

\abst{The IceCube, Pierre Auger and Telescope Array Collaborations have recently reported results on neutral particles (neutrons, photons and neutrinos) which complement the measurements on charged primary cosmic rays at ultra-high energy. The complementarity between these messengers and between their detections are outlined. The current status of their search is reviewed and a cross-correlation analysis between the available results is performed. The expectations for photon and neutrino detections in the near future are also presented.}

\kword{UHECR 2014, multi-messenger, neutron, neutrino, photon}

\begin{document}
\maketitle

\section{Introduction}

The most energetic and violent, but also less understood, astrophysical objects 
in the Universe are expected to produce non-thermal radiation of 
hadronic origin, with an accompanied flux of gamma-rays and
neutrinos. The simultaneous observation of these particles
- the so-called multi-messenger approach - is a key ingredient
for discovering the sources themselves, and for better understanding
the underlying mechanisms responsible for their violent activity.

High-energy multi-messenger astronomy has entered an exciting
era, with the development and operation of new detectors
offering unprecedented opportunities to observe cosmic radiation
in the Universe, and with the successful detection of 
gamma-rays and neutrinos with energies $10^{14}-10^{15}$ eV,
at the lower edge of the high-energy Universe we hope 
to observe in the future.

The aim of this contribution is to review the status of 
high-energy gamma-ray, neutron, and neutrino observations, as well 
as to provide prospects for future observations, with a focus
on the energy range from $10^{15}$ eV up to the highest energy ever
observed in a single particle $\sim 10^{20}$ eV.

\section{Multi-messenger particles}

Identifying the sources of very high-energy (VHE $10^{15}-10^{18}$ eV)
and ultra-high energy (UHE $>10^{18}~{\rm eV}=1~{\rm EeV}$) cosmic rays (CRs) is one of the key problems
in Astroparticle Physics. Being charged particles, cosmic rays are deflected
in the galactic and extragalactic magnetic fields and lose information
about the position of the site where they were produced,
unless they are light nuclei and have energies above a few tens of EeV.
However, even at these energies (more precisely above $\sim 5\times 10^{19}$ eV),
cosmic rays are expected to interact with the Cosmic Microwave Background
(CMB) radiation \cite{G,ZK}, that limits their observability to distances smaller than 50-100 Mpc, representing
a small fraction of the observable volume of the Universe.
Fortunately, high-energy photons and neutrinos are expected to be produced by cosmic rays at/around their sources as well as in their
propagation in the Universe through neutral and charged pion production and their successive decay.

As neutral particles, photons and neutrinos propagate undeflected by magnetic fields, pointing back to their production sites, including the sources of the CRs. 
This makes them suitable candidates to extend astronomical observations
to unprecedented energy ranges.
Moreover, if the sources of CRs are transient, typically involving compact objects
(such as gamma-ray bursts),
an observation of photons and neutrinos (and possibly gravitational waves)
could in fact be the only path that might lead to a full understanding
of the underlying processes.

Photons (up to $\sim$100 TeV) have been observed from a
variety of sources \cite{Holder_review}. Their detection
is already providing important clues on the origin of the very-high energy cosmic
rays, with the hadronic origin of the observed gamma-rays favored with respect
to the leptonic model in a few cases \cite{SNR_Science}. Unfortunately,
the absorption of gamma-rays by electron-positron pair production with the low energy
photons of the extragalactic background radiation (infrared, CMB and radio)
limits their observability to the local Universe. The effect is energy-dependent, 
with photons at energies around $10^{15}$ eV having a horizon distance of order of the
Milky Way size, while UHE photons at 10 EeV are
detectable from distances of $\sim10$ Mpc.

High-energy neutrinos can escape from denser and deeper environments than
photons, and propagate unaltered through the Universe with virtually no limit
to their observation distance regardless of their energy. Neutrinos can provide important information about
the processes taking place in astrophysical engines and could
even reveal the existence of sources opaque to hadrons and
photons, sources that would thus far have remained undetected.
VHE neutrinos of cosmic origin, with energies up to 2 PeV, have recently
been observed, for the first time, with the IceCube experiment \cite{IceCube_PRL14}.
Their detection represents an important milestone and a major step towards
multi-messenger astronomy. Unfortunately, due to the nature of the detected
events, identification of their sources has not been yet possible (cf.\ Sec.\ \ref{sec:NeutrinoSearches}), even if 
natural candidates are the identified sources of TeV photons.

In the UHE range, around 1 EeV and above, neutrinos and gamma-rays
have so far escaped detection by existing experiments and only
upper limits to their fluxes exist. An expected ``guaranteed" source of UHE photons and neutrinos
is the interaction of UHECRs with the CMB. The fluxes are uncertain mainly
due to the unknown energy spectrum and composition of the primary cosmic ray beam, and due to the
evolution with redshift and spatial distribution of the sources. On the other hand,
and thanks to these dependencies, the observation of UHE photons and neutrinos
would provide further hints on the features of UHECR sources \cite{Kotera_GZK}.
Furthermore in the UHE range, top-down models, in which the pions producing
photons and neutrinos arise from the decay or annihilation of exotic particles,
are strongly constrained by searches for UHE photons and neutrinos \cite{Auger_photons,ES,DG}. 

In the following, we give a brief introduction to the observatories involved in this report.

The IceCube Neutrino Observatory (IceCube) has been installed at depths between
1450~m and 2450~m in the Antarctic ice sheet at the geographic South
Pole. It comprises 86 vertical strings with a typical inter-string
spacing of 125~m, each one instrumented with 60 digital optical
modules (DOMs) \cite{Abbasi:2009qf}. 
IceCube is optimized to measure Cherenkov
radiation from charged particles produced in TeV to PeV neutrino
interactions inside or close to the 1~km$^3$ of instrumented ice.

On the surface above the deep ice detector, an array of 81 stations, IceTop \cite{Abbasi:2012zr}, measures charged particles from extensive air showers induced by cosmic rays in the energy range from 300~TeV to $>$
1~EeV. In a limited zenith angle range, IceTop can also be utilized as a
veto against atmospheric muons in the deep ice detector, which
constitute the largest source of background for neutrino searches, and
vice versa for searches for muon-poor showers that constitute
high-energy photon candidates.

The Pierre Auger Observatory (Auger) is located in the Province of Mendoza, Argentina,
at a mean altitude of 1400~m (atmospheric overburden of $\sim875 {\rm~g~cm^{-2}}$) \cite{AugerNew} and sensitive to primary particles of UHE. 
The Observatory is a hybrid system, a combination of a large surface detector array (SD)
and a fluorescence detector (FD). The SD is composed of over
1660 water-Cherenkov stations placed in a triangular grid with nearest neighbors
separated by 1500~m, and spread over an area of $\sim3000~{\rm km^2}$. 
The 27 telescopes of the FD overlook the SD from five sites.
The SD was used to search for UHE neutrinos (cf.\ Sec.\ \ref{sec:NeutrinoSearches}), while
both the SD and FD were used to search for UHE photons (cf.\ Sec.\ \ref{sec:PhotonSearches}). 

The Telescope Array (TA) experiment is located in Millard County, Utah, USA and optimized for detecting UHE particles. It consists of the surface detector with 507 scintillators covering an area of approximately 700~km$^2$ overlooked by 38 fluorescence telescopes located at
three sites~\cite{TASD,TAFD}. The TA fluorescence telescopes operate in hybrid mode with the surface detector
and at the same time allow monocular and stereo reconstruction of the events. 

\section{Results}

\subsection{Neutron searches}
\label{sec.Neutron}

Given the limited lifetime of free neutrons at rest of about 886~s, the mean travel distance is limited for relativistic neutrons to $9.2 \times E$~kpc, where $E$ is the energy of the neutron in EeV. For energies above $2$~EeV most of the galactic disk is in reach for neutron astronomy while the galactic center is already in range of $1$~EeV neutrons. In the energy range of a few EeV Auger, HiRes, and Telescope Array have found their data consistent with a significant component of protons~\cite{Auger:2012wt,Auger:2010,HiRes:2010,TA:2012}. If sources in the galaxy are emitting protons up to the ankle in the energy spectrum, they could show themselves through a flux of primary neutrons produced by pion photo-production and nuclear interactions of the protons near the source.

\begin{figure}[t!]
\includegraphics[width=1\linewidth]{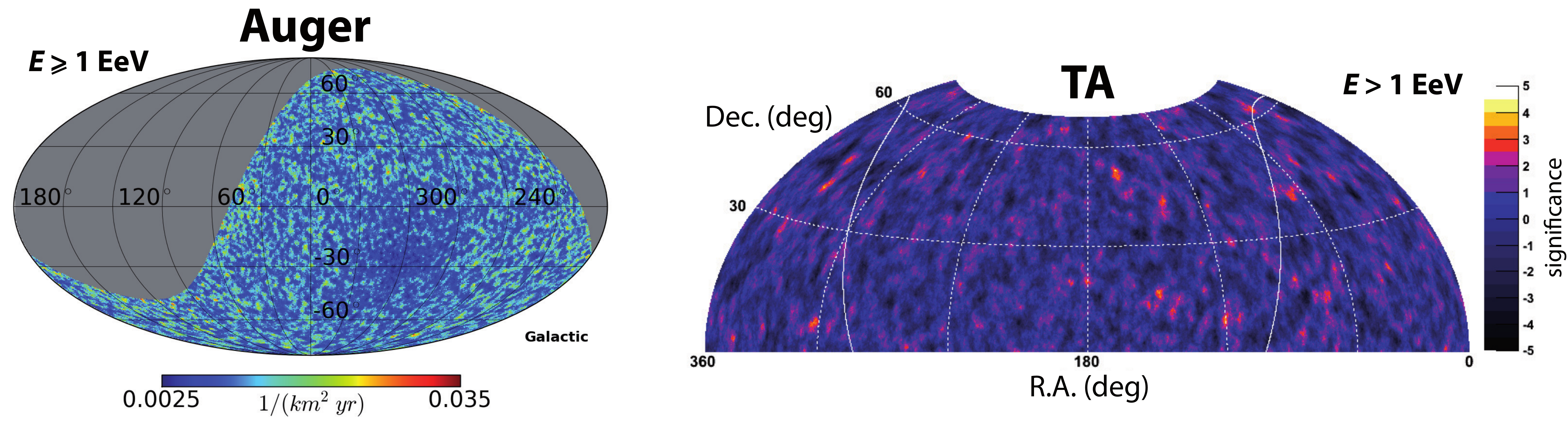}
\caption{\textbf{Left:} Celestial map of the flux upper limit (particles ~km$^{-2}$~yr$^{-1}$) for neutral particles of Auger in galactic coordinates~\cite{AugerNeutron_2012}. \textbf{Right:} Significance map of neutral particles of TA in equatorial coordinates~\cite{Abbasi:2014wza}.}
\label{fig.AugerNeutronBlind}
\end{figure}

Neutron induced air showers are indistinguishable from air showers produced by primary protons. However, since they are not deflected by magnetic fields, 
the signature of a primary neutron flux is an excess of
hadronic air showers from a specific celestial direction. This complements the search for a directional excess of photons discussed in Sec.\ \ref{sec:DirectionalPhotonSearch} where dedicated discrimination techniques are applied. 
Both the Pierre Auger Observatory and Telescope
Array searched for an excess of events in the southern and northern 
hemisphere, respectively. No statistically significant
excess has been found in any small solid angle that would be indicative of
a flux of neutral particles from a discrete
source~\cite{AugerNeutron_2012,Abbasi:2014wza}. In the case of the Pierre
Auger Observatory the search covers a declination band from $-90^\circ$
to $+15^\circ$ in four different energy ranges above 1~EeV. With a total
exposure of 24,880~km$^2$~sr~yr flux upper limits of neutral particles
have been placed in the exposed sky and are displayed in Fig.\
\ref{fig.AugerNeutronBlind} (left). The typical (median) flux upper limit above
1~EeV is 0.0114 neutron~km$^{-2}$~yr$^{-1}$ corresponding to an energy
flux limit of 0.083~eV~cm$^{-2}$~s$^{-1}$. In case of the Telescope Array
experiment, a search is performed in a declination band from $0^\circ$
to $70^\circ$ in four different energy ranges starting from $0.5$~EeV. A significance map derived from TA events is shown in Fig.\
\ref{fig.AugerNeutronBlind} (right). The averaged point
source flux upper limit in the northern sky is 0.07
neutrons~km$^{-2}$~yr$^{-1}$~\cite{Abbasi:2014wza} for the energy greater than $1$~EeV.

To limit the statistical penalty for making many trials (as in the blind search analyses described above) classes of objects of potential sources of neutrons are tested in a separate targeted search~\cite{AugerNeutronTargeted_2014}. In total, nine target sets of candidate sources of a given class are combined in a ``stacking analysis''. The signal should be more significant than that of a single target by itself under the hypothesis that many (or all) of the candidate sources are indeed emitting neutrons. In addition the galactic center and the galactic plane region are considered as additional stand-alone targets. None of the candidate source classes tested revealed compelling evidence for EeV neutrons. Upper limits from the galactic plane constrain models for continuous production of EeV protons in the Galaxy.

\subsection{Neutrino searches}
\label{sec:NeutrinoSearches}

\begin{figure}[hbt]
\centering
\includegraphics[width=8.1cm]{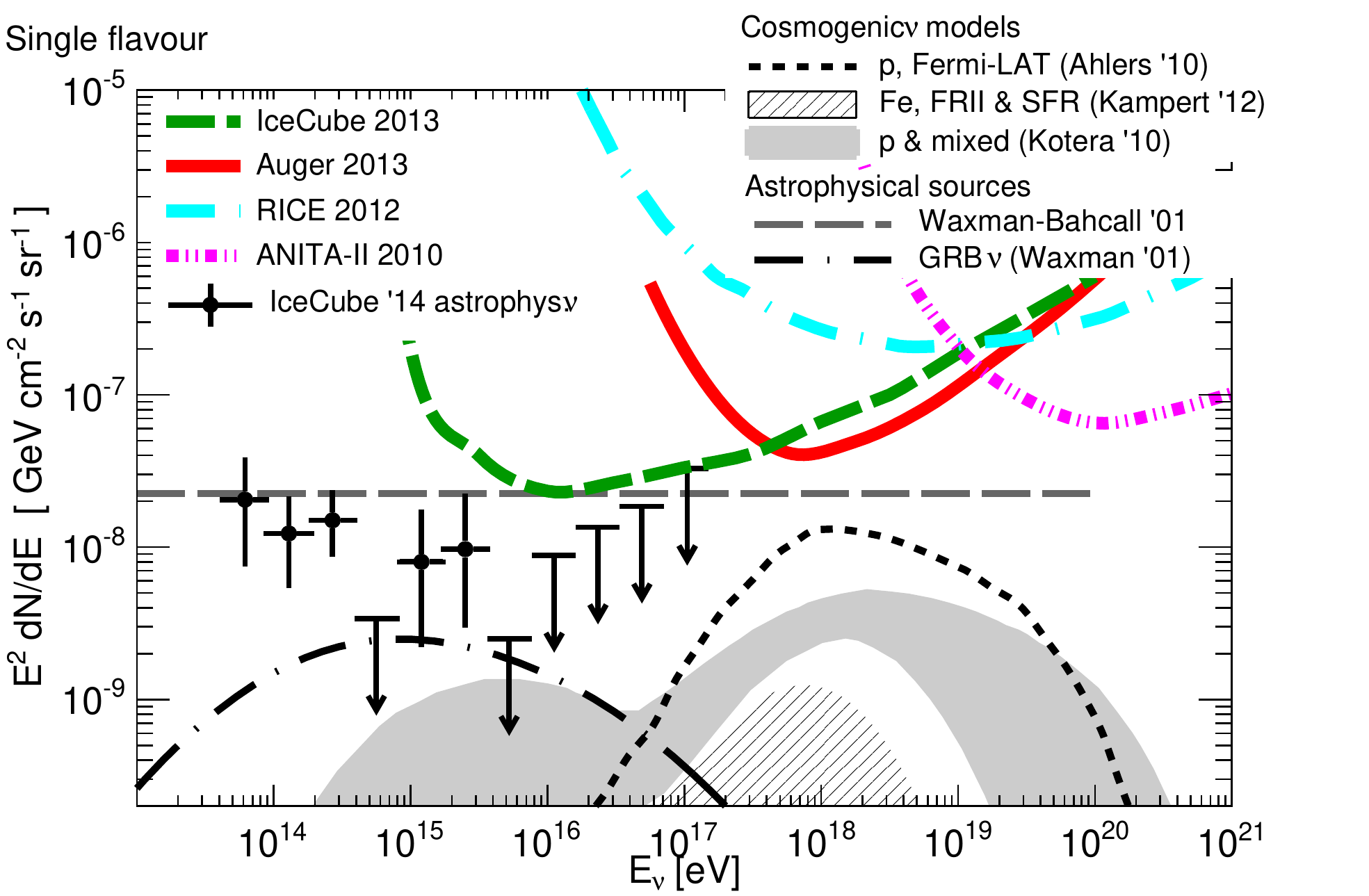}
\includegraphics[width=7.1cm]{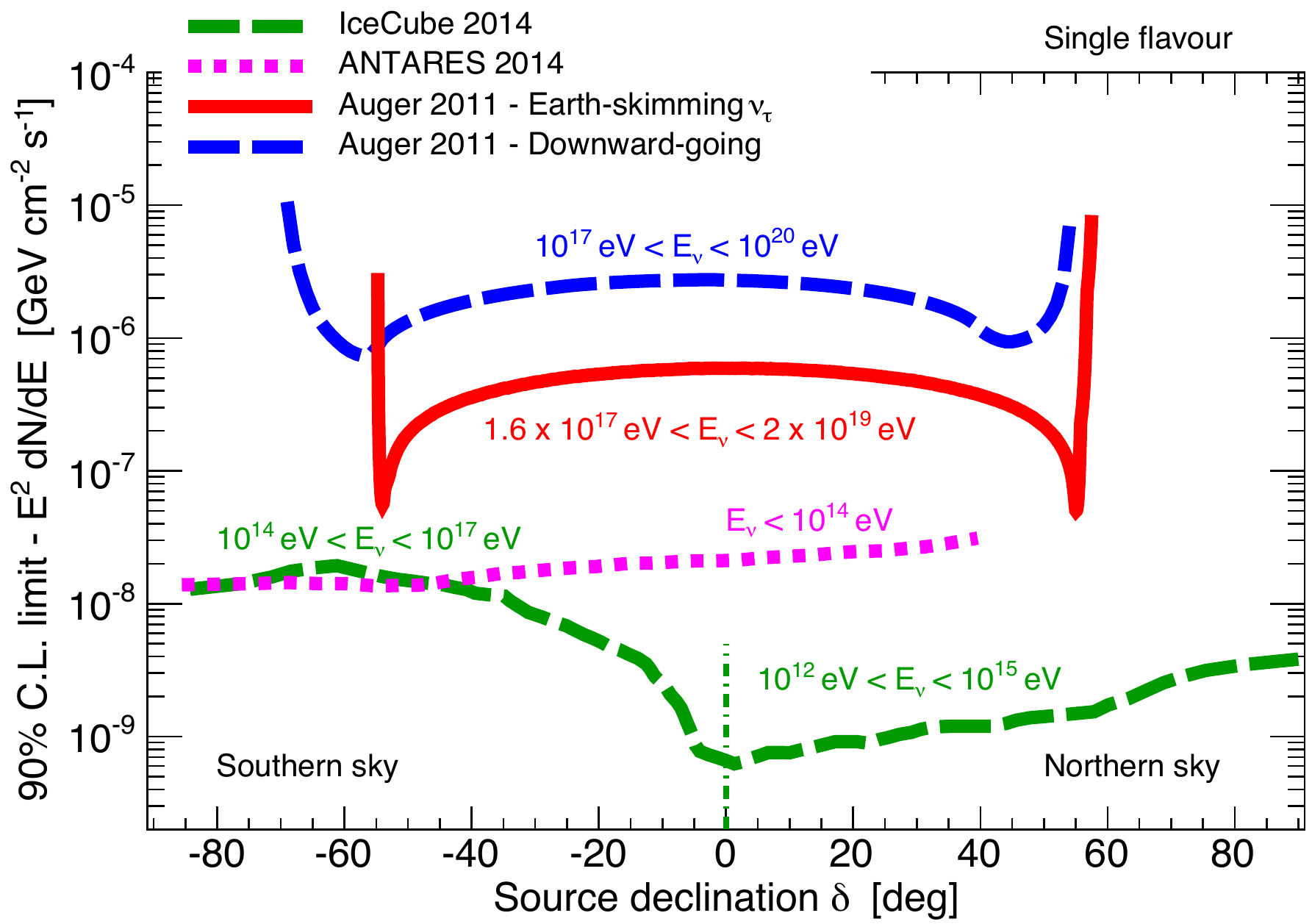}
\vskip -3mm
\caption{ \textbf{Left:} The flux of neutrinos measured by the IceCube
  experiment \cite{IceCube_PRL14} (data points) along with different upper limits from the Pierre Auger Observatory
  \cite{Auger_nus_ICRC13}, ANITAII \cite{ANITAII}, IceCube
  \cite{Aartsen:2013ap} and RICE \cite{RICE_2012} experiments (all limits converted to single flavour and 90\% C.L.\ and shown in bins of width $0.5$ in $\log_{10}E_\nu$ -- except for the RICE limit that is calculated as explained in \cite{RICE_2006}). Also shown are the expected
  neutrino fluxes for several cosmogenic neutrino models
  \cite{Ahlers_GZK,Kampert_GZK,Kotera_GZK} as well as the
  Waxman-Bahcall bound \cite{WB}. \textbf{Right:} Upper limits (at 90$\%$ C.L.) on the
  single-flavour neutrino flux from point-like sources for different
  declinations derived from non-observations in the Auger
  \cite{Abreu:2012ys}, IceCube \cite{Aartsen:2014ad}, and ANTARES
  \cite{Adrian-Martinez:2014ac} experiments. In all cases an unbroken
  $E^{-2}$ energy spectrum is assumed. For southern hemsiphere sources with a high-energy
  cut-off, IceCube is less sensitive.
}
\label{fig:diff_limits}
\end{figure}

In 2013 IceCube reported the detection of two neutrino candidates with
estimated deposited energies of 1~PeV \cite{Aartsen:2013aa}. These
events were discovered in a search for extremely high energy (EHE)
neutrinos, that used simple selection cuts on the total amount of
light and the reconstructed zenith angle to suppress the atmospheric
muon and neutrino backgrounds. At that time the significance for the
two PeV-events not being of atmospheric origin was $2.8 \sigma$
\cite{Aartsen:2013aa}, and the up-to-today most stringent upper limits
on the neutrino flux between 1~PeV and 200~PeV could be derived
\cite{Aartsen:2013ap} as shown in Fig.~\ref{fig:diff_limits} (left). 
Both observed PeV neutrino events appear cascade-like, i.e.~neutrinos interacting
inside the instrumented volume without an outgoing muon track, and the
photon timing distribution in the detector suggests that both events
are down-going, i.e. originating in the southern hemisphere. These
observations prompted the development of a novel search method:
Selecting bright events that start inside the instrumented volume with the
outer layers of DOMs acting as a veto that allows the suppression
of atmospheric muons by $10^{-5}$ \cite{Aartsen:2013ab}. While
atmospheric neutrinos from the northern hemisphere constitute an
irreducible background, atmospheric neutrinos from the southern
hemisphere can be efficiently suppressed by using self-veto
techniques, that involves tagging them via high energy muons originating
from the same air shower \cite{Schonert:2008zl,Gaisser:2014aa}. 
Applying this search method to three years of
IceCube data revealed 37 high-energy neutrino candidates with
deposited energies between 30~TeV and 2~PeV. The null-hypothesis of a
purely atmospheric origin is rejected at $5.7 \sigma$ and, assuming an
unbroken $E^{-2}$ spectrum, the best-fit per-flavour astrophysical flux
is
$E^2 \phi(E) = (0.95 \pm 0.3) \cdot 10^{-8} \, \textrm{GeV} \,
\textrm{cm}^{-2} \, \textrm{s}^{-1} \, \textrm{sr}^{-1}$
\cite{IceCube_PRL14}. Up to today, no source of these astrophysical
neutrino flux could be identified; the data are consistent with
expectations for equal fluxes of all three neutrino flavours \cite{Aartsen:2015aa} and with
isotropic arrival directions \cite{IceCube_PRL14}. A multi-messenger
analysis to correlate the observed neutrinos with UHE charged cosmic
rays measured by the Pierre Auger Observatory and Telescope Array is
in preparation \cite{Christov:2014}.

A search for neutrinos in the energy range around 1 EeV and above 
has been performed with events detected by the surface detector of the Pierre Auger Observatory. 
While protons, heavier nuclei, and photons 
interact shortly after entering the atmosphere, neutrinos can initiate showers 
close to the ground level. At large zenith angles the atmosphere is thick enough 
so that the electromagnetic component of nucleonic cosmic ray-induced showers 
gets absorbed and the shower front at ground level is dominated by muons. 
On the other hand, showers induced by neutrinos deep in the atmosphere have 
a substantial electromagnetic component at the ground.
The current SD is not directly sensitive 
to the muonic and electromagnetic components of the shower separately, nor 
to the depth at which the shower is initiated.  
However the digitization of the signals induced by the passage of shower particles
with Flash Analog to Digital Converters (FADC) with 25~ns time resolution allows one to distinguish 
narrow signals in time (such as those present in inclined showers initiated high in the 
atmosphere), from the broad signals expected in inclined showers initiated close to the ground.
Applying this simple idea we can efficiently detect inclined showers and search for  
(a) Earth-skimming (ES) showers induced by tau neutrinos ($\nu_\tau$)
that travel in the upward direction with respect to the vertical to ground,
and (b) downward-going (DG) showers initiated by any neutrino flavour 
at large zenith angles and that interact 
in the atmosphere close to the SD.
Typically, only Earth-skimming $\nu_\tau$-induced showers with zenith 
angles $90^\circ < \theta < 95^\circ$ and downward-going showers 
with $\theta > 60^\circ$ may be identified with large efficiency. 
The identification of potential neutrino-induced showers 
is based on first selecting those events that arrive in rather inclined directions, 
and then selecting among them those with FADC traces that are spread in time,
indicative of the early stage of development of the shower and 
a clear signature of a deeply interacting neutrino triggering the SD.
Full details of the procedure are given in 
\cite{ES,DG,Auger_nus_ICRC13}.

A novel approach with respect to previous
publications \cite{ES,DG} has been implemented
in the analysis for the first time by combining
searches in the ES and DG channels together using a simple
procedure (see \cite{Auger_nus_ICRC13,Auger_nus_PRD_15}
for details). In this way the exposure to UHE neutrinos is enhanced with respect
to that obtained when each channel is considered separately.
A strong background reduction against showers initiated by 
UHE cosmic rays is possible. The search for UHE neutrinos at the Auger Observatory 
is limited by exposure but not by background. The ES and DG criteria were applied to data between 
1 January 2004 up to 31 December 2012 (excluding data
periods used to train the selections) with no neutrino 
candidates found. 

For the calculation of the exposure of the SD to
UHE neutrinos, the same set of trigger and
neutrino identification conditions that were applied to data are 
also applied to the simulated neutrino showers.
Using the combined exposure  
and assuming a differential neutrino flux $dN(E_\nu)/dE_\nu = k\cdot E_\nu^{-2}$ 
as well as a $\nu_e:\nu_\mu:\nu_\tau=1:1:1$ flavour ratio, 
an upper limit on the value of $k$ at $90\%$ C.L. 
was obtained through a standard procedure \cite{Feldman-Cousins}:
$E^2 dN/dE ~< ~1.3 \times 10^{-8}~{\rm GeV~cm^{-2}~s^{-1}~sr^{-1}}$.
This is below the normalization of the Waxman-Bahcall landmark \cite{WB}. Auger is the first shower array to reach that level of sensitivity (cf.\ page 79, Fig.\ 5 in \cite{Auger_nus_ICRC13}).
The limit is displayed in differential format in 
Fig.~\ref{fig:diff_limits} (left), where the 
Auger limit integrated in bins of width 0.5 in $\log_{10}{E_{\nu}}$ 
is presented, along with those from other experiments \cite{ANITAII,Aartsen:2013ap}. 
As can be seen in Fig.~\ref{fig:diff_limits} (left), the maximum sensitivity of Auger 
is achieved at ${\sim}$EeV neutrino energy where most cosmogenic models
of $\nu$ production also peak (in an $E_\nu^2$ times flux plot). 
The result is finalized with more data until 20
June 2013 in \cite{Auger_nus_PRD_15}.
The Auger limit places strong constraints on cosmogenic $\nu$ models 
that assume a pure primary proton composition injected at the sources
and strong (FRII-type) evolution of the sources \cite{Kampert_GZK}. 
Auger is less sensitive to the cosmogenic neutrino 
models represented by the gray shaded area in Fig.~\ref{fig:diff_limits}
which brackets the lower fluxes predicted under a 
range of assumptions \cite{Kotera_GZK}. 
The same remark applies to models
that assume pure-iron composition at the sources.

No neutrino sources have been identified, yet. While Auger is sensitive to neutrinos with energies larger
than 100~PeV, IceCube can extract a very pure TeV-neutrino sample from
the northern hemisphere which is dominated by atmospheric neutrinos
and a sample of high-energy (100~TeV to 100~PeV) neutrino-candidate
events in the southern hemisphere. In the Auger Observatory
no neutrino candidates were found \cite{Abreu:2012ys}. IceCube has performed dedicated 
searches for spatial clustering of neutrino candidate events but no significant clustering of the arrival directions has been observed \cite{Aartsen:2014ad}. Upper limits on the flux from point-like neutrino
sources are calculated by both experiments and illustrated in Fig.\ \ref{fig:diff_limits} (right). The resulting 90$\%$ C.L.\ upper limits are shown as a function of declination. For the
southern hemisphere the most stringent limits are set by the ANTARES
neutrino telescope in the Mediterranean Sea, especially for sources
with a high-energy cut-off which is expected for Galactic neutrino
sources \cite{Adrian-Martinez:2014ac}.

\subsection{Photon searches}
\label{sec:PhotonSearches}

Given their mostly electromagnetic nature, on average, photon-induced air showers develop deeper in the atmosphere, compared to hadron-induced ones of similar energies. In addition, a significantly smaller muon content, compared to hadron induced showers, is expected. As a consequence, photon induced showers at EeV energies have a smaller signal in the surface detectors, a steeper lateral distribution of secondary particles, a narrower distribution of arrival times of particles in the shower front and a larger delay with respect to a planar shower front approximation. These
differences are further enhanced at energies above 10$^{19}$~eV because of
the Landau-Pomeranchuk-Migdal (LPM) effect~\cite{LPM1,LPM2}.   
The development of the air shower is then slowed and the event by event fluctuations are enhanced
~\cite{dedenko,Billoir}.
At energies above $\sim$50~EeV, photons entering the geomagnetic field 
can convert to electron-positron pairs with a 
probability that depends on the energy of photons and on the component of the
local magnetic field orthogonal to the particle
motion~\cite{preshower1,preshower2}.  A bunch of low-energy
electromagnetic particles, called a ``preshower'', thus enters the
atmosphere and is detected as a single shower developing higher in atmosphere and with a flatter lateral distribution. 

\vspace{0.2cm}
\label{sec:DiffusePhotonSearch}
\emph{Diffuse photon searches at ultra-high energy}

Searches for photons have been performed by the Auger and the TA collaborations. No primary photons have been unambiguously identified so far above the EeV energies and upper limits have been placed on the diffuse photon fraction and integral photon flux by several experiments~\cite{AugerPhotonSD2007,AugerPhotonHy2009,Auger_photons,Abu-Zayyad:2013dii,Auger_photons2011,agasalimits,Yakutsk,HaverahPark}. Above 10$^{19}$~eV, both Auger and TA have performed a search for photon-induced showers using the respective surface detectors. The mass sensitive observables in the two analyses are related to the curvature of the shower front and, in the case of Auger, also to the time spread of particles (see~\cite{Auger_photons,Abu-Zayyad:2013dii} for details).
 At energies lower than 10$^{19}$~eV, upper limits on the integral flux of photons are derived by Auger using events detected with the fluorescence telescopes operating in hybrid mode. These events benefit from the low energy threshold of the direct observation of the depth at which the energy deposit reaches its maximum (\Xm) with the FD and of  complementary informations from SD~\cite{Auger_photons2011}. 
 An analysis based on the observation of \Xm with hybrid events is in progress within the TA collaboration. 
  
\begin{figure}[!t]
\hspace{-2mm}
\vspace{0mm}
\hspace{-0mm}
\begin{minipage}[c]{0.5\textwidth}
  \includegraphics[width=1.0\textwidth]{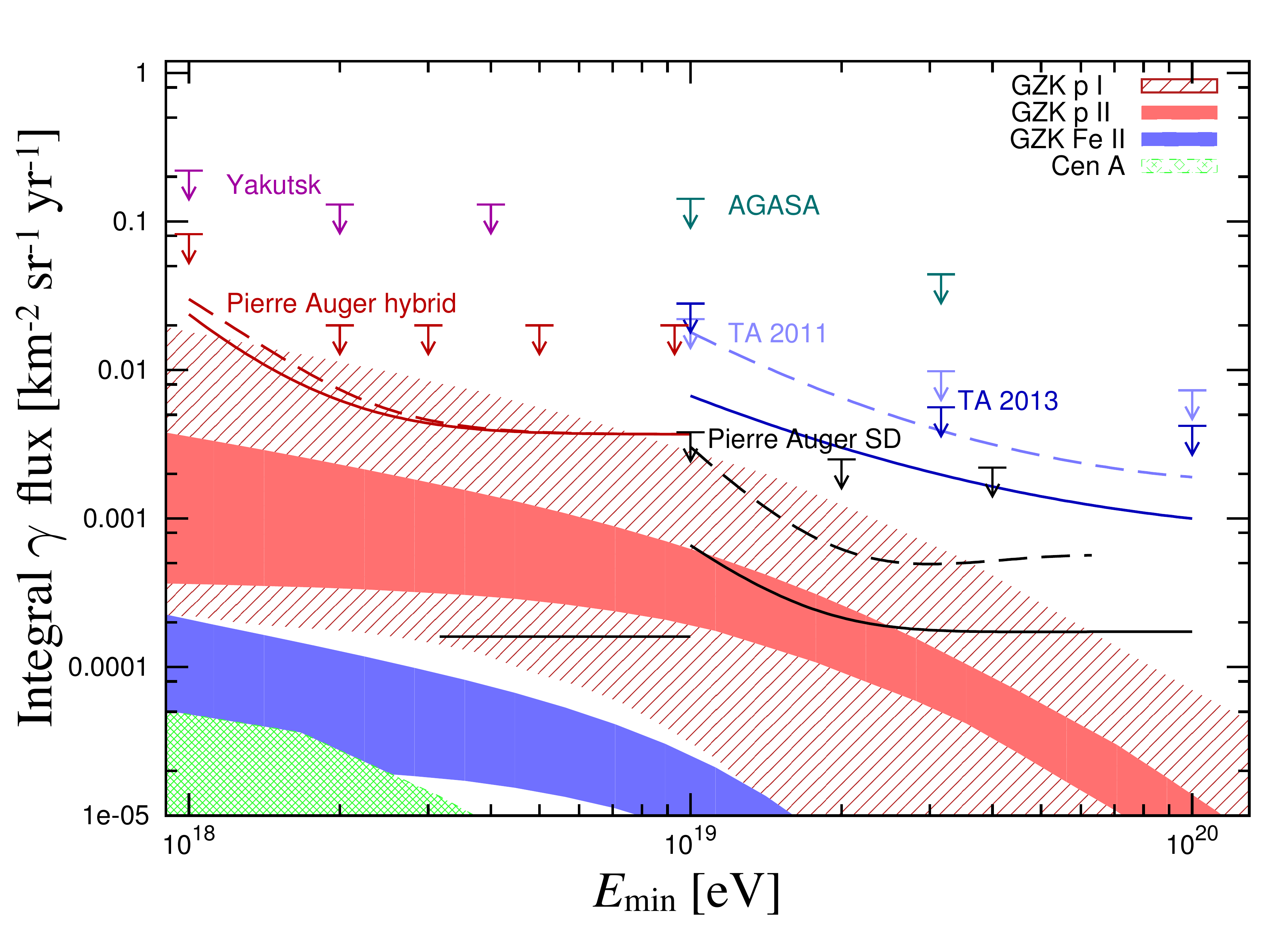}
\end{minipage}
\begin{minipage}[c]{0.6\textwidth}
\includegraphics[width=0.82\textwidth]{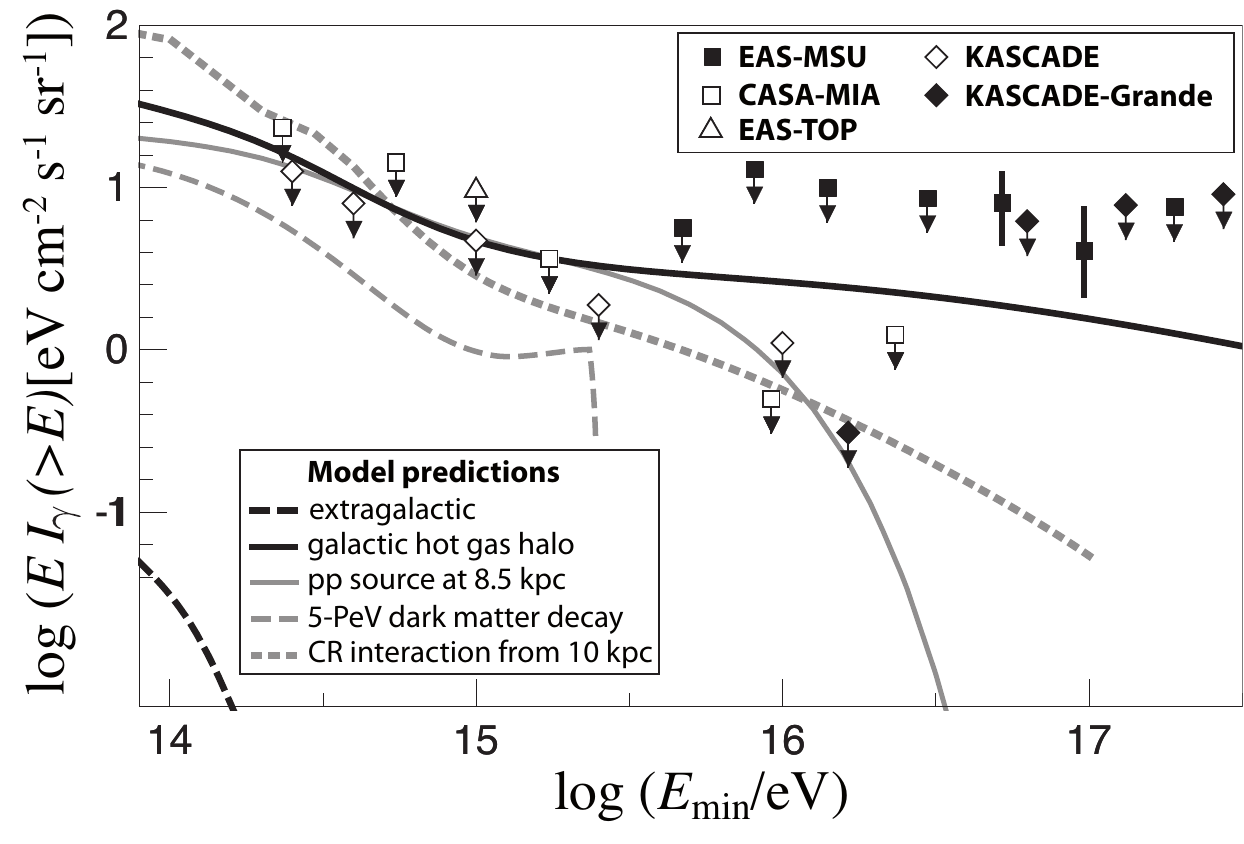}
\end{minipage}
  \caption{\textbf{Left:} 95\% C.L.\ upper limits on the integral UHE photon flux derived by
Auger with the hybrid~\cite{Auger_photons2011} and SD~\cite{Auger_photons} detectors
and TA SD detector~\cite{Abu-Zayyad:2013dii},
compared to the results of AGASA~\cite{agasalimits} and Yakutsk~\cite{Yakutsk}. The shaded regions give the predictions for the GZK photon flux models from~\cite{gelmini} assuming protons at the source (red shaded), from~\cite{Hooper} (red and blue, under the assumption of proton and iron acceleration), and for the case of a single source as Centaurus A~\cite{CenA} (green). Sensitivity expectations of the year 2020 for Auger and TA are indicated by dashed and solid lines, cf.\ Sec.\ \ref{sec:Persp} for more details. \textbf{Right:} 
PeV to EeV constraints on the diffuse photon flux.
Fluxes are obtained assuming isotropic emission; data from
EAS-TOP (90\% C.L.) \cite{EAS-TOP}, KASCADE (C.L.\ not quoted)
\cite{KASCADE, MMWG2012}, KASCADE-Grande (90\% C.L.)~\cite{KASCADE-Photons}, CASA-MIA (90\% C.L.) \cite{CASA-MIA} and
EAS-MSU (95\% C.L.) \cite{EAS-MSU-flux1}. Systematic errors for all these results, mostly related to simulations of the background contamination, are of order $\pm 50\%$. Lines represent some model predictions constrained by the
IceCube measured neutrino flux: extragalactic \cite{KT},
Galactic hot gas halo \cite{KT}, $pp$ source at a fixed
distance of 8.5~kpc \cite{Kohta}, 5-PeV dark matter decay
\cite{Kohta}, cosmic-ray interactions with interstellar
matter at a fixed distance of 10~kpc \cite{Walter}.
}
\label{fig:UL}
 \end{figure}

The current upper limits on the integral photon flux are shown in Fig.~\ref{fig:UL} as a function of the energy reconstructed assuming that the primary particle is a photon. In fact, while for hybrid events the energy is directly measured by FD in a calorimetric way, for SD events the energy estimator is the signal at a reference distance $r_{\rm opt}$ specific to each experimental layout and the SD detector. The difference in the development between the bulk of air showers and potential rare photon-induced events results in incorrect determination of the energy for primary photons \cite{BilloirEnergy, KRT}. This effect is taken into account and all modern constraints on the flux of primary UHE photons use the appropriate energy scale for photon primaries.
The tighter limits found by Auger, compared to TA, reflect the differences in the detectors and in the exposure of the two experiments: (i) the exposure of TA is about a factor 7 smaller than that of Auger; (ii) compared to scintillators,  the water-Cherenkov stations are more sensitive to muons, so that they detect weaker signals from the muon-poor photon induced showers, enhancing the photon/hadron separation capabilities. 
Moreover the two experiments, located in different hemispheres, have quite different and complementary fields of view, so a direct comparison should be taken with care. 

The current results disfavor the exotic models for the origin of UHECR described in~\cite{gelmini,SHDM2} over a wide energy range while the region of expected GZK photon fluxes in the most optimistic scenario can be explored by Auger in the very near future, cf.\ Fig.\ \ref{fig:UL} (left).  

\vspace{0.2cm}
\label{sec:DiffusePhotonSearch}
\emph{Diffuse photon searches at TeV -- PeV energy}

Following the multi-messenger paradigm that cosmic rays, photons and neutrinos are produced in the same hadronic processes in the sources, the recent discovery of an excess of high-energy neutrinos over the atmospheric background (cf.\ Sec.\ \ref{sec:NeutrinoSearches}) has attracted considerable interest~\cite{Gupta-gamma, A-exposure, Kohta, Kohta1, Walter, KT} to searches for a diffuse photon flux at PeV energies. Experimental constraints on the diffuse flux of sub-PeV to sub-EeV
photons are presented in Fig.~\ref{fig:UL} (right) together with predictions from models. Note that EAS-MSU reported a claim of detection of cosmic gamma rays above PeV energies~\cite{EAS-MSU-flux1}. 
However, the KASCADE-Grande experiment did not find any signal in the same energy range with similar sensitivity~\cite{KASCADE-Photons}, cf.\ Fig.\ \ref{fig:UL} (right).

The IceCube Observatory also is sensitive to PeV photons by looking for muon-poor showers in the deep ice-detector. A search has been performed with the 40 string configuration in the declination region  $-90^\circ < \delta < -60^\circ$ \cite{Aartsen:2012uq}.
No correlation of photon candidates with the Galactic plane was found and an upper limit on the photon fraction of $1.2 \cdot 10^{-3}$ was set in the energy range from 1.2 to 6~PeV. Further, no clustering of photon candidate events was observed in a search for point-like sources in the complete field of view. A similar analysis with the full IceCube detector is in preparation.

\label{sec:DirectionalPhotonSearch}
\vspace{0.35cm}
\emph{Directional photon searches}

Similar to the search for a directional excess of neutrons discussed in
Sec.\ \ref{sec.Neutron}, ultra-high energy photons, produced in vicinity
of a potential source, can be detected by an accumulation of events from a
specific target direction. Since photon induced air-showers exhibit
specific characteristics, the sensitivity for detection
can additionally be improved by selecting only ``photon-like'' events.
Compared to EeV neutrons, the detectable volume is large enough to
encompass in addition to the Milky Way, the Local Group of galaxies, and
possibly Centaurus A, given an attenuation length of about 4.5~Mpc at EeV
energies~\cite{Risse2007,DeAngelis:2013jna,EleCa}.

\begin{figure}[t!]
\begin{center}
\includegraphics[width=0.65\linewidth]{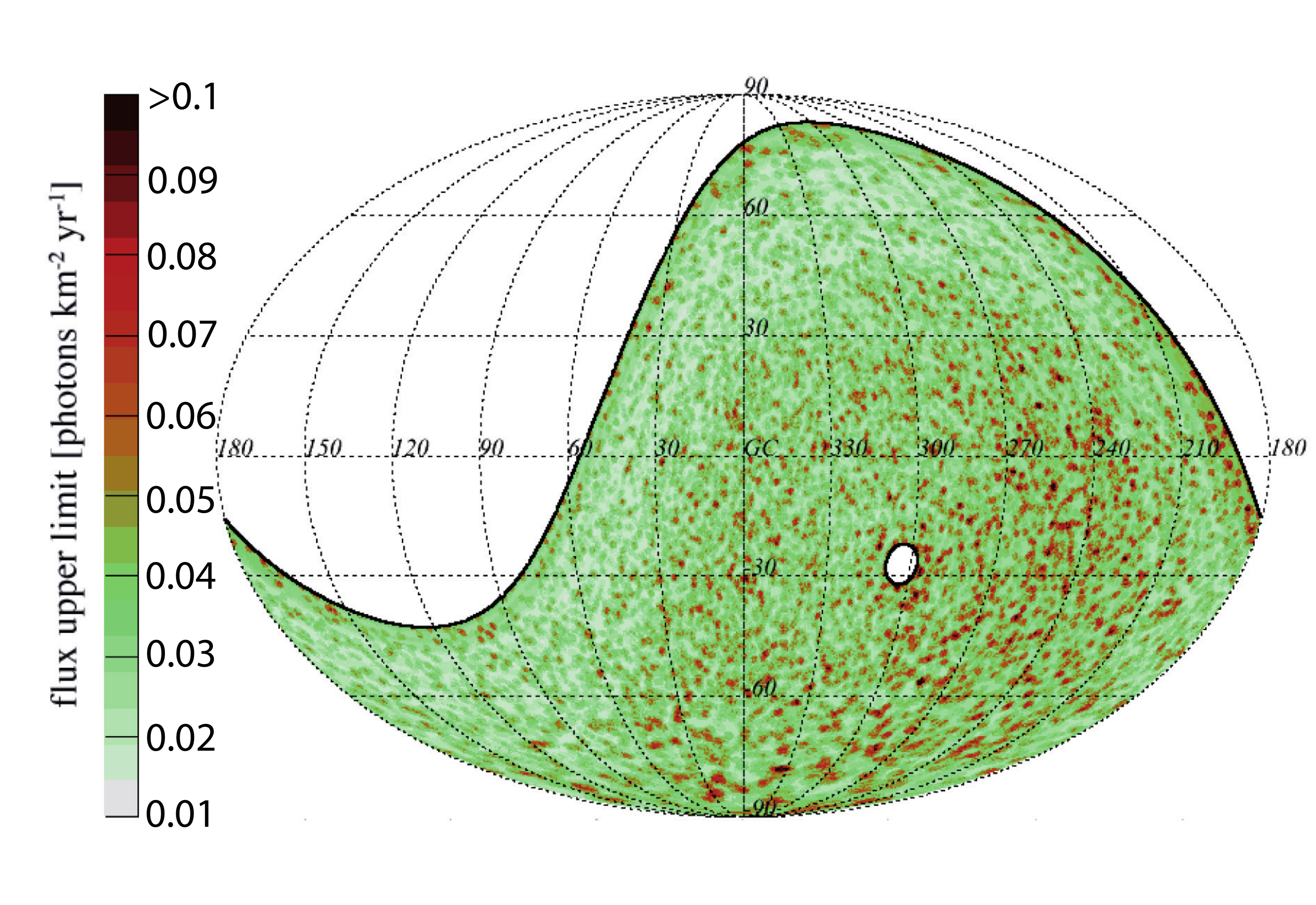}
\caption{ 
Celestial map of photon flux upper
limits in photons~km$^{-2}$~yr$^{-1}$ illustrated in Galactic coordinates.
\cite{Aab:2014bha}}
\label{fig.AugerPhotonBlind}
\end{center}
\end{figure}

Data from the Pierre Auger Observatory was used to search for an excess of
photon-like events from 526,200 target directions (target separation $\sim
0.3^\circ$) between a declination range of $-85^{\circ}$ and $+20^{\circ}$
in the energy range of $10^{17.3}$~eV up to $10^{18.5}$~eV
\cite{Aab:2014bha}. To select photon-like events, five photon-sensitive
observables from the surface detector array as well as from the
fluorescence detector are combined in a multivariate analysis using
boosted decision trees~\cite{Breiman1984,Schapire1990}. 
Details on the photon-like events selection and on the statistical methods used to derive the $p$-value (local probability that the data is in agreement with
a uniform distribution) and the upper limits on the photon flux are given in~\cite{Aab:2014bha}. 
No statistical evidence for any target direction has been found.  The
mean value for the upper limit at 90\% C.L.\ is 0.035~photons~km$^{-2}$~yr$^{-1}$, with a maximum of
0.14~photons~km$^{-2}$~yr$^{-1}$ as shown in Fig.\
\ref{fig.AugerPhotonBlind}. Those values correspond to an energy flux of
0.06~eV~cm$^{-2}$~s$^{-1}$ and 0.25~eV~cm$^{-2}$~s$^{-1}$, respectively,
assuming an $E^{-2}$ energy spectrum. 

The limits are of considerable astrophysical interest. By extrapolating
measured energy fluxes at TeV energies (e.g.\
\cite{Hinton:2009zz,HESSCite}) assuming a spectral index of $E^{-2}$ this
flux would have been detected with more than $5\sigma$ significance, even
after penalizing for the large number of trials. Furthermore, making
conservative assumptions for median exposure targets 
one can exclude a photon flux greater than 1.44~eV~cm$^{-2}$~s$^{-1}$ with
$5\sigma$ significance. The null result from this search does not mean
that the sources of EeV cosmic rays are extragalactic as they might be
produced for example by transient sources where the Earth was not exposed during the
collection of data or continuous sources in the Galaxy which emit in jets
that do not point towards the Earth.

The directional photon search is also
underway in TA.

\section{Cross-correlation analysis between Auger and TA}

As we have already pointed out, Auger and TA are complementary in the photon
search both by their fields of view and by techniques. However, in the declination band of $-20^\circ < {\rm dec} < +20^\circ$ both experiments have considerable exposure. Neither Auger nor TA have seen an excess of photon-like events,
from any direction, statistically significant in a full-sky search,
penalized for the fact that no particular direction is a priory expected
to host a source. There exist, however, local excesses of photon-like
events in the Auger data and a number of photon candidates in the TA data.
The directions to these excesses should be uncorrelated if they are just
statistical fluctuations while a directional coincidence between local
excesses of photon-like events seen by two independent experiments may
indicate, if significant, the existence of a real UHE gamma-ray source.
The purpose of the joint Auger$+$TA study, whose preliminary results are
presented here for the first time, is to search for such possible
correlations by looking for statistically significant excesses of Auger
photon-like events from the directions of TA photon candidates. For this
analysis, we use preliminary arrival directions of 9 TA surface detector
photon-like events with energies $E_{\gamma}>10$~EeV and angular resolution of $1.4^\circ$ and 8 TA hybrid
photon-like events with $E>2$~EeV and angular resolution of $0.9^\circ$ in the common field of view described above. The bulk of events in the Auger data set have considerably
lower energies ($10^{17.3} < E ~[\log (E/{\rm eV})] < 10^{18.5}$), so this kind of analysis may constrain
sources emitting photons in both energy ranges only.

To calculate the chance probability of the correlation of photon candidate
events from the Pierre Auger Observatory and the Telescope Array the
analysis is divided into several steps. First, a combined $p$-value is
calculated from TA candidate directions. For $n$ photon candidate events
in the field of view of Auger with
directions $(\alpha_1,\delta_1)$, $(\alpha_2,\delta_2)$, ...
,$(\alpha_n,\delta_n)$, $P^{\rm combined}$ is defined as
\begin{equation}
P^{\rm combined} = \prod_n p^{\rm Auger} (\alpha_n,\delta_n)~,
\label{eqn_combined}
\end{equation}
where $p^{\rm Auger}$ is the weighted average $p$-value from a specific direction. This value is obtained by weighting the actual $p$-value from \cite{Aab:2014bha} with a von Mises-Fisher distribution (vMF)~\cite{Mises-Fisher} which incorporates in its concentration parameter the angular resolution of TA events. 

To determine if the resulting $P^{\rm combined}$ is significantly small,
mock $P^{\rm random}$-values are obtained from MC simulations. These are calculated in
the same way as in Eqn.\ \ref{eqn_combined} but using $n$ randomly
distributed arrival directions according to the acceptance of the TA
detector. The last step is repeated $\sim 1,000,000$ times and the
resulting distribution of $P^{\rm random}$ values is compared with $P^{\rm
combined}$ by calculating the chance probability $p^{\rm chance} (P^{\rm random} \leq P^{\rm combined})$.

\begin{figure}[bt]
\centering
\includegraphics[width=15.0cm]{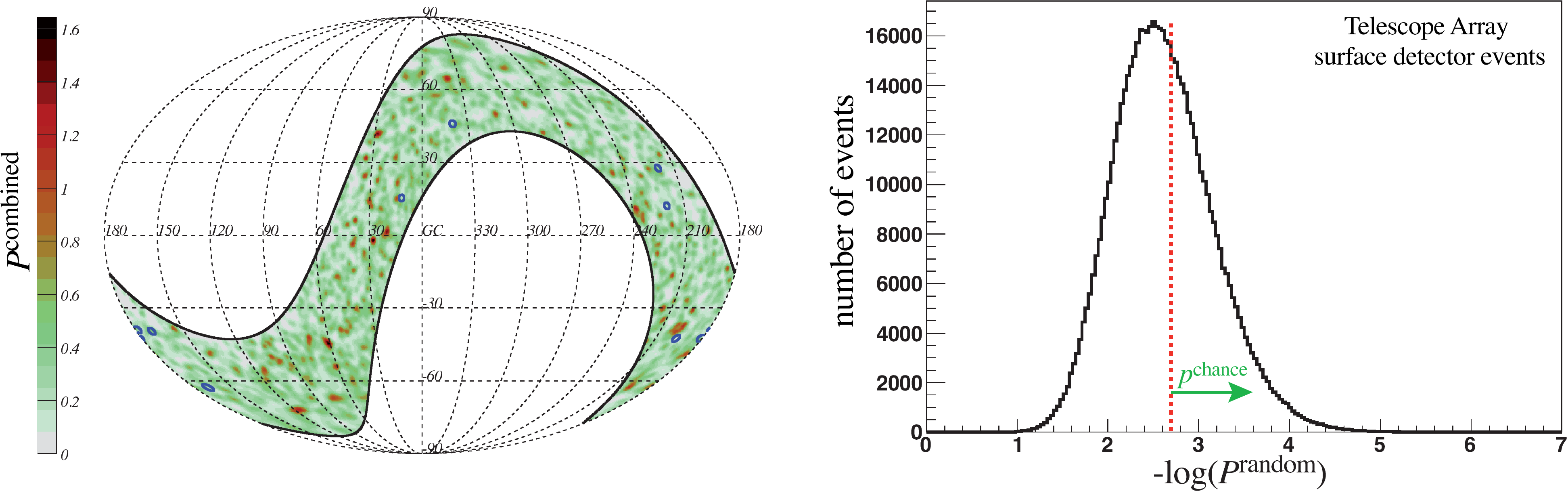}
\vskip -3mm
\caption{\textbf{Left:} Sky distribution of $P^{\rm combined}$ values in galactic coordinates. Only the common observation band is shown. Blue circles indicate the position of 9 TA surface detector photon-like events with a radius according their angular resolution. \textbf{Right:} Distribution of $P^{\rm random}$ values obtained from MC simulations. The red dashed line indicates the observed combined $p$-value of $-\log (P^{\rm combined}) = 2.71$.}
\label{fig:CrossCorr}
\end{figure}

The results of this analysis, using as an example the photon candidate events detected by the TA surface detector, are shown in Fig.\ \ref{fig:CrossCorr}. The corresponding chance probability is $p_{\rm SD}^{\rm chance} = 41\%$. Using photon-like events from the TA hybrid detector a chance probability of $p_{\rm Hy}^{\rm chance} = 68\%$ is obtained. Both analyses indicate that there is no statistically significant excess of photon candidate events when combining photon-like events from the Pierre Auger Observatory and Telescope Array in the commonly observed declination band. 
We must remark here that such result is not surprising since the photon-like events in the Auger data sample as well as the TA candidate events used as targets in the analysis, are mainly dominated by the nuclear background. 

\section{Summary and perspectives}
\label{sec:Persp}
In this contribution, we have reported about the status of the search for neutral primary cosmic rays with particular emphasis on the search for photons and neutrinos with Auger, Telescope Array and IceCube observatories.  Whereas IceCube has observed a flux of astrophysical neutrinos, no UHE photons and neutrinos have been identified by Auger and Telescope Array. A flux of such primary particles is expected as result of the UHECRs interaction with the CMB. A multi-messenger approach would provide strong constraints on the astrophysical scenarios  as photon and neutrino fluxes are sensitive to different features of the source environment and of the cosmic-ray propagation. 
As already discussed, given the large exposure of the Auger Observatory, 
cosmogenic neutrino flux models are already being constrained, with detection
of the first UHE cosmogenic neutrinos in reach in the next few years 
if the primary cosmic ray flux is dominated by protons. The Auger Observatory 
is most sensitive in an energy region complementary to the one where IceCube has
its maximum detection capabilities. Also, planned experiments detecting radio 
waves in the MHz-GHz frequency range produced through the Askaryan 
effect in neutrino-induced showers (such as the Askaryan Radio Array -- ARA \cite{ARA} 
and ARIANNA \cite{ARIANNA}) have competing expected sensitivities to EeV neutrinos from cosmological origin. 
Last but not least, the analysis of data collected at the recent ANITA-III antenna-payload
balloon flight is in progress.

To identify the sources of the astrophysical neutrinos an extension to
IceCube, IceCube-Gen2, is currently being designed
\cite{Aartsen:2014ak} in the southern hemisphere. With an instrument volume of about 10~km$^{3}$
it will be able to gather sufficient statistics with improved angular
resolution to identify neutrino point sources or rule out many source
classes. Further, it will allow us to precisely measure the energy
spectrum and flavour composition of the diffuse neutrino flux. In the northern hemisphere, two km$^3$-scale projects, KM3NeT~\cite{KM3NET} and Baikal-GVD~\cite{Baikal}, are under active development.

Interesting outcomes are expected in the near future also for photon primaries at ultra-high energy, as Auger will be able to explore the region of expected cosmogenic photon fluxes in the most optimistic astrophysical scenarios. 
The sensitivity to the photon flux expected in 2020 with Auger and TA is shown in Fig.~\ref{fig:UL} left (dashed and solid lines). 
The dashed lines are the extrapolations of the current analyses: compared to the estimates
in~\cite{MMWG2012}, they take into account the detection of background events 
as derived from the current knowledge of the hadron-shower
contamination. A significant boost of these sensitivities (solid lines) is possible with
the upgrades of the Auger detectors~\cite{KampertUtah2014} and the extensions of the TA array~\cite{Sagawa:2014qja}. In the case of TA, the expected limits are derived assuming an improved
analysis and three years of extension of the surface array by a factor
4. For Auger, an ideal scenario is shown, which includes new trigger algorithms, recently installed over the SD array, and the possible upgrade of the Observatory with muon detectors. Since a full analysis procedure is not yet established, the extrapolated limits are derived under the assumptions of 50\% efficiency in photon selection and no detected photon and background events.
Moreover, these results can have implications for fundamental physics
since the predicted flux of GZK photons can be affected by Lorentz
invariance violation (see for example~\cite{LIV,LIV2}) or by possible
photon-axion conversion~\cite{RaffeltStodolsky, axions} during photon
propagation.

Combining the results of present and future instruments in a
multi-messenger approach offers a chance to get insights on the cosmic ray puzzle
within the next decade.


\begin{thebibliography}{99}

\bibitem{G}
 K.~Greisen,
  %``End to the cosmic ray spectrum?,''
  Phys.\ Rev.\ Lett.\  {\bf 16} (1966) 748.
  %%CITATION = PRLTA,16,748;%%
  %2076 citations counted in INSPIRE as of 16 Feb 2015

\bibitem{ZK}
G.~T.~Zatsepin and V.~A.~Kuzmin,
  %``Upper limit of the spectrum of cosmic rays,''
  JETP Lett.\  {\bf 4} (1966) 78
   [Pisma Zh.\ Eksp.\ Teor.\ Fiz.\  {\bf 4} (1966) 114].
  %%CITATION = JTPLA,4,78;%%
  %1561 citations counted in INSPIRE as of 16 Feb 2015

\bibitem{Holder_review}
J. Holder, Astropart. Phys. {\bf 39-40} (2012) 61.

\bibitem{SNR_Science}
M. Ackermann et al.\ (Fermi-LAT Collaboration),
Science {\bf 339} (2013) 807.

\bibitem{IceCube_PRL14}
M.~G. Aartsen et al.\ (IceCube Collaboration),
Phys.\ Rev.\ Lett.\ {\bf 113} (2014) 101101.

\bibitem{Kotera_GZK}
D. Allard, K. Kotera and A. Olinto,
JCAP {\bf 10} (2010) 013.

\bibitem{Auger_photons}
J.\ Abraham et al. (Pierre Auger Collaboration),
Astropart.\ Phys.\ {\bf 29} (2008) 243.

\bibitem{ES}
J. Abraham et al.\ 
Phys.\ Rev.\ Lett.\ {\bf 100} (2008) 211101;
Phys.\ Rev.\ D {\bf 79} (2009) 102001;
P.\ Abreu et al.\ 
Astrophys.\ J.\ Lett.\ {\bf 755} (2012) L4.

\bibitem{DG}
P. Abreu et al.\ (Pierre Auger Collaboration),
Phys.\ Rev.\ D {\bf 84} (2011) 122005;
J.~L. Navarro, PhD Thesis, Univ. Granada, Spain (2012).

\bibitem{Abbasi:2009qf}
R.~Abbasi et~al.\ (IceCube Collaboration),
%``The {IceCube} data acquisition system: Signal capture, digitization, and timestamping,''
  Nucl.\ Instrum.\ Meth.\ A
  {\bf 601} (2009) 294.

\bibitem{Abbasi:2012zr}
R.~Abbasi et~al.\ (IceCube Collaboration),
%%``{IceTop}: The surface component of {IceCube},''
Nucl.\ Instrum.\ Meth.\ A {\bf 700} (2013) 188.

\bibitem{AugerNew}
A.\ Aab et al.\ (Pierre Auger Collaboration),
Nucl.\ Instrum.\ Meth.\ A {\bf 798} (2015) 172.  %arXiv:1502.01323 [astro-ph.IM]

\bibitem{TASD} 
  T.~Abu-Zayyad et al.\ (Telescope Array Collaboration),
  %``The surface detector array of the Telescope Array experiment,''
  Nucl.\ Instrum.\ Meth.\ A {\bf 689}  (2012) 87.

\bibitem{TAFD} 
  H.~Tokuno et al.\ (Telescope Array Collaboration),
  %``New air fluorescence detectors employed in the Telescope Array experiment,''
  Nucl.\ Instrum.\ Meth.\ A {\bf 676} (2012) 54.

\bibitem{Auger:2012wt}
  P.~Abreu et al.\ (Pierre Auger Collaboration),
  %``Measurement of the proton-air cross-section at $\sqrt{s}=57$ TeV with the Pierre Auger Observatory,''
  Phys.\ Rev.\ Lett.\  {\bf 109} (2012) 062002.
  %%CITATION = ARXIV:1208.1520;%%

\bibitem{Auger:2010}
  J.~Abraham et al.\  (Pierre Auger Collaboration),
  %``Measurement of the Depth of Maximum of Extensive Air Showers above 1018 eV,''
  Phys.\ Rev.\ Lett.\  {\bf 104} (2010) 091101.

\bibitem{HiRes:2010}
  R.~U.~Abbasi et al.\  (HiRes Collaboration),
  %``Indications of Proton-Dominated Cosmic-Ray Composition above 1.6 EeV,''
  Phys.\ Rev.\ Lett.\  {\bf 104} (2010) 0161101.

\bibitem{TA:2012}
  C.~C.~H.~Jui for the Telescope Array Collaboration,
  %``Cosmic Ray in the Northern Hemisphere: Results from the Telescope Array Experiment,''
  J.\ Phys.: Conf.\ Ser.\  {\bf 404} (2012) 012037.

\bibitem{AugerNeutron_2012}
  P.~Abreu et al.\ (Pierre Auger Collaboration),
  %``A SEARCH FOR POINT SOURCES OF EeV NEUTRONS,''
  Astrophys.\ J.\  {\bf 760} (2012) 148.

\bibitem{Abbasi:2014wza}
  R.~U.~Abbasi et al.\  (Telescope Array Collaboration),
  %``A Northern Sky Survey for Point-Like Sources of EeV Neutral Particles with the Telescope Array Experiment,''
  arXiv:1407.6145 [astro-ph.HE].
  %%CITATION = ARXIV:1407.6145;%%

\bibitem{AugerNeutronTargeted_2014}
  A.~Aab et al.\ (Pierre Auger Collaboration),
  %``A TARGETED SEARCH FOR POINT SOURCES OF EeV NEUTRONS,''
  Astrophys.\ J.\ Lett.\ {\bf 789} (2014) L34.

\bibitem{Aartsen:2013aa}
M.~G. Aartsen et~al.\ (IceCube Collaboration),
%``First observation of {PeV-energy} neutrinos with {IceCube},''
Phys.\ Rev.\ Lett. {\bf 111} (2013) 021103.

\bibitem{Aartsen:2013ap}
M.~G. Aartsen et~al.\ (IceCube Collaboration),
%``Probing the origin of cosmic rays with extremely high energy neutrinos using the {IceCube} observatory,''
Phys.\ Rev.\ D {\bf 88} (2013) 112008.

\bibitem{Aartsen:2013ab}
M.~G. Aartsen et~al.\ (IceCube Collaboration),
%``Evidence for high-energy extraterrestrial neutrinos at the {IceCube} detector,''
 Science {\bf 342} (2013) 1242856.

\bibitem{Schonert:2008zl}
S.~Sch{\"o}nert, T.~K. Gaisser and E.~Resconi et~al.,
%``Vetoing atmospheric neutrinos in a high energy neutrino telescope,''
 Phys.\ Rev.\ D {\bf 79} (2009) 043009.

\bibitem{Gaisser:2014aa}
T.~K. Gaisser, K.~Jero, A.~Karle et~al.,
%``A generalized self-veto probability for atmospheric neutrinos,''
  Phys.\ Rev.\ D {\bf 90} (2014) 023009.

\bibitem{Aartsen:2015aa}
M.~G. Aartsen et al.\ (IceCube Collaboration),
%``Flavor ratio of astrophysical neutrinos above {35 TeV} in
%icecube,''
Phys.\ Rev.\ Lett.\ {\bf 114} (2015) 171102.% arXiv:1502.03376 [astro-ph.HE].

\bibitem{Christov:2014}
A.~Christov, G.~Golup and M.~Rameez,
%``Towards a joint analysis of data from the {IceCube} neutrino telescope, the {Pierre Auger Observatory} and  {Telescope Array},''
this conference proceedings.

\bibitem{Auger_nus_ICRC13}
P.\ Pieroni for the Auger Collaboration,
Proc.\  33rd ICRC 2013, Rio de Janeiro, arXiv:1307.5059 [astro-ph].

\bibitem{Auger_nus_PRD_15}
A. Aab et al.\ (Pierre Auger Collaboration),
Phys.\ Rev.\ D {\bf 91} (2015) 092008.

\bibitem{Feldman-Cousins}
G.~J. Feldman, R.\ D.\ Cousins, Phys.\ Rev.\ D {\bf 57} (1998) 3873;
J.\ Conrad et al., Phys.\ Rev.\ D {\bf 67} (2003) 012002.

\bibitem{WB}
E. Waxman and J.~N. Bahcall,
Phys. Rev. D {\bf 59} (1998) 023002;
Phys. Rev. D {\bf 64} (2001) 023002.

\bibitem{ANITAII}
P.~W. Gorham et al. (ANITA Collaboration),
%Phys. Rev. D {\bf 82} (2010) 022004;
Phys.\ Rev.\ D {\bf 85} (2012) 049901(E).

\bibitem{RICE_2012}
I.\ Kravchenko et al.\ (RICE Collaboration),
Phys.\ Rev.\ D {\bf 73} (2012) 082002.

\bibitem{RICE_2006}
I.\ Kravchenko et al.\ (RICE Collaboration),
Phys.\ Rev.\ D {\bf 65} (2006) 062004.

\bibitem{Kampert_GZK}
K.-H. Kampert and M. Unger, 
Astropart. Phys. {\bf 35} (2012) 660. 

\bibitem{Ahlers_GZK}
M.\ Ahlers et al., Astropart.\ Phys.\ {\bf 34} (2010) 106.

\bibitem{Abreu:2012ys}
P.~Abreu et al.\ (Pierre Auger Collaboration),
Astrophys.\ J.\ Lett.\ {\bf 755} (2012)  L4.

\bibitem{Aartsen:2014ad}
M.~G. Aartsen et al.\ (IceCube Collaboration),
Astrophys.\ J.\ {\bf 796} (2014) 109.

\bibitem{Adrian-Martinez:2014ac}
S.~Adri{\'a}n-Martinez et al.\ (ANTARES Collaboration),
Astrophys.\ J.\ Lett.\ {\bf 786} (2014)  L5.

\bibitem{LPM1}
  L.~D. Landau, I.~Ya. Pomeranchuk, Dokl.\ Akad.\ Nauk,
  SSSR {\bf 92} (1953) 535.

\bibitem{LPM2}
A.~B.\ Migdal, Phys.\ Rev.\ {\bf 103} (1956) 1811.

\bibitem{dedenko}
L.~G. Dedenko
et al.,
Proc.\ 17th ICRC, Paris {\bf 7} (1981) 159.

\bibitem{Billoir}
X.\ Bertou, P.\ Billoir and S.\ Dagoret-Campagne,
Astropart.\ Phys. {\bf 14} (2000) 121.

\bibitem{preshower1}
  T. Erber, Rev.\ Mod.\ Phys. {\bf 38} (1966) 626.

\bibitem{preshower2}
  B. McBreen and C.~J. Lambert, Phys.\ Rev.\ D {\bf 24} (1981) 2536.

\bibitem{AugerPhotonSD2007}
J.\ Abraham et al.\ (Pierre Auger Collaboration),
Astropart.\ Phys. {\bf 27} (2007) 155.

\bibitem{AugerPhotonHy2009}
J.\ Abraham et al.\ (Pierre Auger Collaboration),
Astropart.\ Phys. {\bf 31} (2009) 399.

\bibitem{Abu-Zayyad:2013dii} 
  T.~Abu-Zayyad et al.\  (Telescope Array Collaboration),
  %``Upper limit on the flux of photons with energies above $10^{19}$  eV using the Telescope Array surface detector,''
  Phys.\ Rev.\ D {\bf 88} (2013) 112005.

\bibitem{Auger_photons2011}
M.\ Settimo for the Pierre Auger Collaboration,
Proc. 32nd ICRC, Beijing, {\bf 2}: 51 (2011).% arXiv:1107.4805.

\bibitem{agasalimits}
  K. Shinozaki et al., Astrophys. J. {\bf 571} (2002) L117.

\bibitem{Yakutsk}
  A. Glushkov et al., Phys.\ Rev.\ D {\bf 82} (2010) 041101.

\bibitem{HaverahPark}
M.~Ave, J.~A.~Hinton, R.~A.~Vazquez, A.~A.~Watson and E.~Zas,
  %``New constraints from Haverah Park data on the photon and iron fluxes of UHE cosmic rays,''
  Phys.\ Rev.\ Lett.\  {\bf 85} (2000) 2244.

\bibitem{BilloirEnergy}
 P.\ Billoir, C.\ Roucelle, and J.~C.\ Hamilton, astro-ph/0701583.

\bibitem{KRT}
O.~E.~Kalashev, G.~I.~Rubtsov and S.~V.~Troitsky,
  %``Sensitivity of cosmic-ray experiments to ultra-high-energy photons: reconstruction of the spectrum and limits on the superheavy dark matter,''
  Phys.\ Rev.\ D {\bf 80} (2009) 103006.
  %  [arXiv:0812.1020 [astro-ph]].
  %%CITATION = ARXIV:0812.1020;%%
  %11 citations counted in INSPIRE as of 16 Feb 2015

\bibitem{gelmini}
  G.~Gelmini, O.~Kalashev, D.~Semikoz, J.\ Exp.\ Theor.\ Phys.
  {\bf 106} (2008) 1061.

\bibitem{SHDM2}
  J. Ellis et al., Phys.\ Rev.\ D,
 {\bf 74} (2006) 115003.

\bibitem{Hooper}
  D.\ Hooper, A.\ M.\ Taylor and S.\ Sarkar, Astropart.\ Phys.,
  {\bf 34} (2011) 340.

\bibitem{CenA}
  M. Kachelriess, S. Ostapchenko and R. Tomas, Publ. Astron.\ Soc.\ Aust.,
  {\bf 27} (2010) 482.

\bibitem{EAS-TOP}
M.~Aglietta, B.~Alessandro, P.~Antoni et al.\  (EAS-TOP Collaboration),
  %``A Limit to the rate of ultrahigh-energy gamma-rays in the primary cosmic radiation,''
  Astropart.\ Phys.\  {\bf 6} (1996) 71.
  %%CITATION = APHYE,6,71;%%
  %12 citations counted in INSPIRE as of 17 May 2013

\bibitem{KASCADE}
%G.~Schatz, F.~Fessler, T.~Antoni
G.~Schatz
%, W.~D.~Apel, F.~Badea, K.~Bekk, A.~Bercuci and H.~Blumer
et al.\ (KASCADE collaboration),
  %``Search for extremely high energy gamma rays with the KASCADE experiment,''
Proc.~28th ICRC, Tsukuba {\bf 4} (2003) 2293.
  %%CITATION = FZKA-6890I;%%

\bibitem{MMWG2012}
  J. Alvarez-Muniz, M. Risse, G.I. Rubtsov, B.T. Stokes for the Pierre Auger, Telescope Array and Yakutsk Collaborations,
  EPJ Web of Conferences {\bf 53} (2013) 01009.

\bibitem{CASA-MIA}
M.~C.~Chantell et al.\ (CASA-MIA Collaboration),
%``Limits on the isotropic diffuse flux of ultrahigh-energy gamma radiation,''
  Phys.\ Rev.\ Lett.\  {\bf 79} (1997) 1805.
%  [astro-ph/9705246].
  %%CITATION = ASTRO-PH/9705246;%%
  %32 citations counted in INSPIRE as of 17 May 2013

\bibitem{EAS-MSU-flux1}
Yu.~A.~Fomin et al.,
%, N.~N.~Kalmykov, G.~V.~Kulikov, V.~P.~Sulakov and
%S.~V.~Troitsky,
JETP Lett.\ {\bf 100} (2014) 797.
%arXiv:1410.2599 [astro-ph.HE].
%%CITATION = ARXIV:1410.2599;%%

\bibitem{KT}
O.~E.~Kalashev and S.~V.~Troitsky,
  %``IceCube astrophysical neutrinos without a spectral cutoff and (10^15-10^17) eV cosmic gamma radiation,''
JETP Letters {\bf 100} (2014) 865.
%  arXiv:1410.2600 [astro-ph.HE].
  %%CITATION = ARXIV:1410.2600;%%
  %5 citations counted in INSPIRE as of 17 Feb 2015

\bibitem{Kohta}
M.~Ahlers and K.~Murase,
  %``Probing the Galactic Origin of the IceCube Excess with Gamma-Rays,''
  Phys.\ Rev.\ D {\bf 90} (2014) 023010.
%  [arXiv:1309.4077 [astro-ph.HE]].
  %%CITATION = ARXIV:1309.4077;%%

\bibitem{Walter}
J.~C.~Joshi, W.~Winter and N.~Gupta,
  %``How Many of the Observed Neutrino Events Can Be Described by Cosmic Ray Interactions in the Milky Way?,''
  Mon.\ Not.\ R.\ Astron.\ Soc.\ {\bf 439} (2014) 3414.
  %%CITATION = ARXIV:1310.5123;%%
  %8 citations counted in INSPIRE as of 11 Oct 2014

\bibitem{Gupta-gamma}
N.~Gupta,
  %``Galactic PeV Neutrinos,''
  Astropart.\ Phys.\  {\bf 48} (2013) 75.
%  [arXiv:1305.4123 [astro-ph.HE]].
  %%CITATION = ARXIV:1305.4123;%%

\bibitem{A-exposure}
L.~A.~Anchordoqui, H.~Goldberg, M.~H.~Lynch et al.,
  %``Pinning down the cosmic ray source mechanism with new IceCube data,''
  Phys.\ Rev.\ D {\bf 89} (2014) 083003.
%  [arXiv:1306.5021 [astro-ph.HE]].
  %%CITATION = ARXIV:1306.5021;%%
  %37 citations counted in INSPIRE as of 01 Aug 2014

\bibitem{Kohta1}
 K.~Murase, M.~Ahlers and B.~C.~Lacki,
  %``Testing the Hadronuclear Origin of PeV Neutrinos Observed with IceCube,''
  Phys.\ Rev.\ D {\bf 88} (2013) 121301.
%  [arXiv:1306.3417 [astro-ph.HE]].
  %%CITATION = ARXIV:1306.3417;%%
  %45 citations counted in INSPIRE as of 11 Oct 2014

\bibitem{KASCADE-Photons}
D.\ Kang for the KASCADE-Grande Collaboration,
Proc. 24th ECRS, Kiel, (2014) in press.

\bibitem{Aartsen:2012uq}
M.~G. Aartsen et~al.\ (IceCube Collaboration),
%``Search for galactic {PeV} gamma rays with the {IceCube} neutrino observatory,''
Phys.\ Rev.\ D {\bf 87} (2013) 062002.

\bibitem{Risse2007}
 M. Risse and P. Homola, Mod.\ Phys.\ Lett.\ A,
{\bf 22} (2007) 749. %astro-ph/0702632.

\bibitem{DeAngelis:2013jna}
  A.~De Angelis, G.~Galanti and M.~Roncadelli,
  %``Transparency of the Universe to gamma rays,''
  Mon.\ Not.\ R.\ Astron.\ Soc.\ {\bf 432} (2013) 3245.
%  [arXiv:1302.6460 [astro-ph.HE]].
  %%CITATION = ARXIV:1302.6460;%%
  %2 citations counted in INSPIRE as of 25 Nov 2013

  \bibitem{EleCa}
 M.~Settimo and M.~De Domenico,
 Proc.\ 33rd ICRC, Rio de Janeiro, (2013), arXiv:1307.3739 [astro-ph.HE].

\bibitem{Aab:2014bha}
  A.~Aab et al.\  (Pierre Auger Collaboration),
  %``A search for point sources of EeV photons,''
  Astrophys.\ J.\  {\bf 789} (2014) 160.
 % [arXiv:1406.2912 [astro-ph.HE]].

\bibitem{Breiman1984}
  L.~Breiman, J.~Friedman, R.~Olshen et al.,
  %``Classification and Regression Trees,''
  Monterey (CA), Wadsworth and Brooks (1984).

\bibitem{Schapire1990}
  R.~E.~Schapire,
  %``The Strength of Weak Learnability,''
  Mach.\ Learn.\ {\bf 5} (1990) 197.

\bibitem{Hinton:2009zz}
  J.~A.~Hinton and W.~Hofmann,
  %``Teraelectronvolt astronomy,''
  Annu.\ Rev.\ Astro.\ Astrophys.\ {\bf 47} (2009) 523.
  %%CITATION = ARXIV:1006.5210;%%
  %62 citations counted in INSPIRE as of 14 May 2013

\bibitem{HESSCite}
  A.~Abramowski et al.\ (H.E.S.S.~Collaboration),
  A\&A {\bf 528} (2011) A143.

\bibitem{Mises-Fisher}
N.~I..~Fisher, T.~Lewis and B.~J.~J. Embleton,
\textit{Statistical analysis of spherical data}, Cambridge Univ.\ Press, 1993.

\bibitem{ARA}
P. Allison et al.\
(ARA Collaboration),
Astropart.\ Phys.\ {\bf 35} (2012) 457.
%and arXiv:1404.5285 [astro-ph].

\bibitem{ARIANNA}
S.~W.\ Barwick et al.\
(ARIANNA Collaboration), Astropart.\ Phys.\ {\bf 70} (2015) 12. % arXiv:1410.7352 [astro-ph].

\bibitem{Aartsen:2014ak}
M.~G. Aartsen et al.\ (IceCube-Gen2 Collaboration),
arXiv:1412.5106 [astro-ph.HE].

\bibitem{KM3NET}
  A.~Margiotta for the KM3NeT Collaboration,
  J.\ Instrum.\ {\bf 9} (2014) C04020.

\bibitem{Baikal}
  A.~V.~Avrorin et al.,
  %``Current status of the BAIKAL-GVD project,''
  Nucl.\ Instrum.\ Meth.\ A {\bf 725} (2013) 23.
  %%CITATION = NUIMA,A725,23;%%

\bibitem{KampertUtah2014}
  K.-H.\ Kampert, ``Auger upgrade program'', this conference proceedings.

\bibitem{Sagawa:2014qja} 
  H.~Sagawa\ (Telescope Array Collaboration),
  %``Highlights from the Telescope Array Experiment,''
  Braz.\ J.\ Phys.\  {\bf 44} (2014) 589.
  %%CITATION = BJPHE,44,589;%%

\bibitem{LIV}
  M. Galaverni and G. Sigl, Phys.\ Rev.\ Lett. {\bf 100} (2008) 021102.

\bibitem{LIV2} 
  G.~Rubtsov, P.~Satunin and S.~Sibiryakov, Phys.\ Rev.\ D {\bf 89} (2014) 123011.

\bibitem{RaffeltStodolsky}
 G.~Raffelt and L.~Stodolsky,
  %``Mixing of the Photon with Low Mass Particles,''
  Phys.\ Rev.\ D {\bf 37} (1988) 1237.
  %%CITATION = PHRVA,D37,1237;%%
  %348 citations counted in INSPIRE as of 16 Feb 2015

\bibitem{axions}
%E. Gabrielli, K. Huitu, S. Roy, Phys. Rev. D {\bf 74} (2006) 073002.
M.~Fairbairn, T.~Rashba and S.~V.~Troitsky,
  %``Photon-axion mixing and ultra-high-energy cosmic rays from BL Lac type objects - Shining light through the Universe,''
  Phys.\ Rev.\ D {\bf 84} (2011) 125019.
%  [arXiv:0901.4085 [astro-ph.HE]].
  %%CITATION = ARXIV:0901.4085;%%
  %42 citations counted in INSPIRE as of 16 Feb 2015

\end{thebibliography}
\end{document}